\documentclass[twocolumn]{aastex631}
\usepackage{graphicx}   
\usepackage{lipsum} 
\usepackage{amsmath}

\usepackage{color}

\def\hmpc{h^{-1}{\rm Mpc}}

\def\hkpc{h^{-1}\, {\rm kpc}}

\def\hmsun{{h^{-1} M_{\odot}}}
\def\msun{\, M_{\odot}}

\def\astrid{\texttt{ASTRID} }
\shorttitle{Star-Forming Wakes and Runaway BHs}
\shortauthors{N.Chen et al.}
\graphicspath{{./}{figures/}}

\begin{document}


\title{Fly-by galaxy encounters with multiple black holes produce star-forming linear wakes}

\author{Nianyi Chen}
\affiliation{McWilliams Center for Cosmology, Department of Physics, Carnegie Mellon University, Pittsburgh, PA 15213}

\author{Patrick LaChance}
\affiliation{McWilliams Center for Cosmology, Department of Physics, Carnegie Mellon University, Pittsburgh, PA 15213}

\author{Yueying Ni}
\affiliation{Center for Astrophysics $\vert$ Harvard \& Smithsonian, Cambridge, MA 02138, US}

\author{Tiziana Di Matteo}
\affiliation{McWilliams Center for Cosmology, Department of Physics, Carnegie Mellon University, Pittsburgh, PA 15213}

\author{Rupert Croft}
\affiliation{McWilliams Center for Cosmology, Department of Physics, Carnegie Mellon University, Pittsburgh, PA 15213}

\author {Priyamvada Natarajan}
\affiliation{Black Hole Initiative, Harvard University, Cambridge, MA 02138, USA}

\affiliation{Department of Astronomy, Yale University, New Haven, CT 06511, USA}

\affiliation{Department of Physics, Yale University, New Haven, CT 06520, USA}

\author{Simeon Bird}
\affiliation{Department of Physics \& Astronomy, University of California, Riverside, 900 University Ave., Riverside, CA 92521, US}


\correspondingauthor{Nianyi Chen}
\email{nianyic@andrew.cmu.edu}

\begin{abstract}
We look for simulated star-forming linear wakes such as the one recently discovered by \cite{vanDokkum2023} in the cosmological hydrodynamical simulation \texttt{ASTRID}.
Amongst the runaway black holes in \texttt{ASTRID}, none are able to produce clear star-forming wakes. Meanwhile, fly-by encounters, typically involving a compact galaxy (with a central black hole) and a star-forming galaxy (with a duo of black holes) reproduce remarkably well many of the key properties (its length and linearity; recent star formation, etc.) of the observed star-forming linear feature. We predict the feature to persist for approximately 100 Myr in such a system and hence constitute a rare event. The feature contains a partly stripped galaxy (with $M_{\rm gal}=10^9 \sim 10^{10}M_\odot$) and a dual BH system ($M_{\rm BH}=10^5 \sim 10^7\,M_\odot$) in its brightest  knot. X-ray emission from AGN in the knot should be detectable in such systems. After $100\sim 200\,{\rm Myrs}$ from the first fly-by, the galaxies merge leaving behind a triple black hole system in an (still) actively star-forming early-type remnant of mass $\sim 5\times 10^{10}\,M_\odot$. Follow-up JWST observations may be key for revealing the nature of these linear features by potentially detecting the older stellar populations constituting the bright knot. Confirmation of such detections may therefore help discriminate a fly-by encounter from a massive BH wake to reveal the origin of such features.


\end{abstract}


\section{Introduction}
\label{section:introduction}

Given that nearly all galaxies host central black holes, it is likely that during complex dynamical mergers involving multiple galaxies, one or more of the BHs get ejected. Evidence from observations for such anticipated three-body interactions has been sparse till recently. 
New observations have reported serendipitous discoveries of extended linear star-forming features with unclear origins that likely involve multiple BH interactions and accompanying triggered star formation. \cite{vanDokkum2023} (VD23 hereafter) reports the discovery of a thin stellar streak extending $\sim 60\,{\rm kpc}$ from a nearby compact galaxy at $z\sim 1$. Based on the length and linearity of the feature, as well as the emission lines associated with dense star-forming gas, the authors interpret it as the stellar wake induced by the passage of a massive black hole (MBH) kicked out from the compact galaxy during a 3-body encounter of the MBHs. 
In another recent paper, \cite{Zaritsky2023} reports finding an even longer, collimated galactic tail $380\,{\rm kpc}$ in length, also associated with recent and ongoing star formation. 
Having discussed the challenges of a more commonplace origin for this feature such as ram pressure or tidal stripping, these authors also propose that this feature is likely created by a three-body encounter between three interacting galaxies. 
Other observed cases of potential runaway BHs include the one detected in a peculiar source in the COSMOS survey at $z=0.359$ (CXOCJ100043.1+020637 (CID-42)) that presents as two compact optical sources embedded in the same galaxy with a large velocity offset \citep{Civano+2010}. 

In this work, we focus on linear features with accompanying star formation activity. 
If the observed linear features were indeed produced by a runaway BH in a compact star cluster, or even a runaway galaxy itself, they open up a new channel for finding ejected stars and MBHs undergoing complex dynamical interaction. 
However, the runaway BH explanation itself also faces theoretical challenges: it is not clear, for example, whether a BH with $M_{\rm BH}\sim 10^7\,M_\odot$ can trigger such a high level of star formation, and leave a trail of stars weighing almost half the mass of the original galaxy ($> 10^9\, M_\odot$ as per VD23). 
On the other hand, it is also theoretically challenging to produce bright stellar wakes that are so thin and straight, without having some ejected material associated with them. 
Further clarification on the origin of these exotic systems demands work from the theoretical side, and the first step would be to identify theoretical counterparts of these observed systems in cosmological simulations, a path we pursue here, so that we can better investigate their origins.

To this end, our work presents a first attempt to query and investigate the origin of extended linear stellar wakes in cosmological simulations and their possible association with runaway MBHs. 
We use the large volume \texttt{ASTRID} simulation with a large galaxy population \citep{Bird-Asterix, Ni-Asterix}, allowing us to find the very rare occurrences of extended star-forming systems.  
Moreover, the updated BH dynamics model \citep[e.g.][]{Chen2021} produces a vast number of wandering MBHs \citep{DiMatteo2022} and even potential runaway MBHs among MBH triplets \citep[][]{Hoffman2023}, allowing us to look for connections between stellar wakes and MBH tuples or runaway BHs.

Our letter is organized as follows: in Section \ref{section2:Method} we first introduce the \texttt{ASTRID} simulation, and the selection criteria we used to find runaway BHs in the simulation as well as the theoretical counterparts of the currently observed linear features. 
In Section \ref{section:Results}, we present systems from the simulation that morphologically resemble the observed linear star-forming features. 
We then examine their origin, evolution, and detailed properties of the galaxies and BHs associated with such systems.
\section{Method}
\label{section2:Method}

\begin{figure*}
    \centering
    \includegraphics[width=0.98\textwidth]{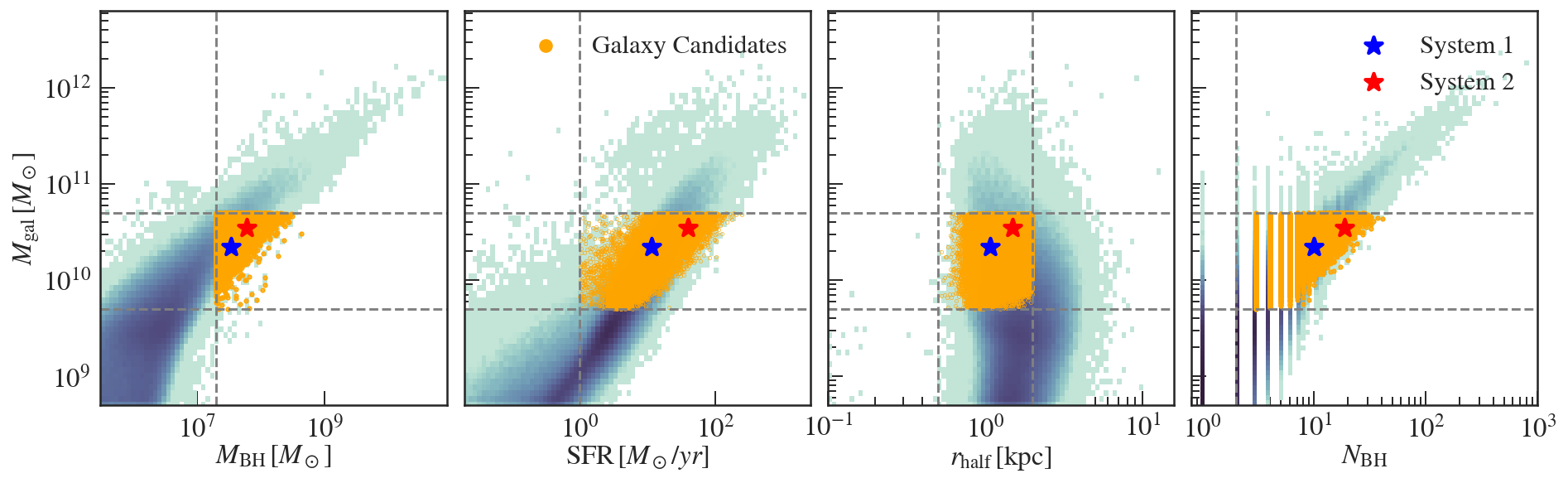}
    \caption{Target galaxies for the search of runaway BHs and linear stellar features (\textit{orange points}), plotted on top of the distribution of all galaxies with $M_{\rm gal} > 5\times 10^8\,M_\odot$ in \texttt{ASTRID} at $z=2$. 
    The two systems with the most prominent linear star-forming features are shown as stars ($z\sim 2$ system in \textit{blue}, $z\sim 1$ system in \textit{red}).
    We focus our search among the star-forming (SFR$ > 1\,M_\odot/{\rm yr}$), compact ($r_{\rm half} < 2 {\rm kpc}$) galaxies with masses $5\times 10^9\,\msun < M_{\rm gal} < 5\times 10^{10}\,\msun$ that contains at least 3 BHs with a central $M_{\rm BH} > 10^7\,\msun$ BH, as indicated by the grey dashed lines. }
    \label{fig:z2population}
\end{figure*}

\subsection{Simulations}
\astrid is a cosmological hydrodynamical simulation performed using a new version of the Smoothed Particle Hydrodynamics code \texttt{MP-Gadget}. The simulation evolves a cube of $250 \hmpc$ per side with $2\times5500^3$ initial tracer particles comprising dark matter and baryons, and has currently reached $z=1.3$. \astrid has a dark matter particle mass resolution of $M_{\rm DM} = 6.7 \times 10^6 \hmsun$ and $M_{\rm gas} = 1.3 \times 10^6 \hmsun$ in the initial conditions. The gravitational softening length is $\epsilon_{\rm g} = 1.5 \hkpc$ for both DM and gas particles. The simulation includes a full-physics sub-grid treatment for modeling galaxy formation, SMBHs and their associated supernova and AGN (Active Galactic Nuclei) feedback, as well as inhomogeneous hydrogen and helium reionization. We refer the readers to the introductory papers \cite{Bird-Asterix,Ni-Asterix} for detailed descriptions of physical models deployed in the simulation. 

Here we briefly summarize BH modeling, which is most relevant to our current investigation of runaway MBHs. BHs are seeded in haloes with $M_{\rm halo,FOF} > 5 \times 10^9 \hmsun$ and $M_{\rm *,FOF} > 2 \times 10^6 \hmsun$, with seed masses stochastically drawn between $3\times10^{4} \hmsun$ and $3\times10^{5} \hmsun$, motivated by the direct collapse scenario proposed in \cite{LodatoPN2007}. 
The gas accretion rate onto the black hole is estimated via a Bondi-Hoyle-Lyttleton-like prescription \citep{DSH2005}. 
The black hole radiates with a bolometric luminosity $L_{\rm bol}$ proportional to the accretion rate $\dot{M}_\bullet$, with a mass-to-energy conversion efficiency $\eta=0.1$ in an accretion disk according to \cite{Shakura1973}. 
5\% of the radiated energy is coupled to the surrounding gas as the AGN feedback. 
Dynamics of the black holes are modeled with a sub-grid dynamical friction model \citep{Tremmel2015,Chen2021} to replace the original implementation that directly repositioned BHs to the local minimum of the potential. 
This gives well-defined black hole trajectories and velocities.
Per this implementation, two black holes can merge if their separation is within two times the gravitational softening length $2\epsilon_g$, once their kinetic energy is dissipated by dynamical friction and they are gravitationally bound.

\subsection{Galaxy Population and Candidate Selection}
\label{sec:selection}



Galaxies in the simulation are identified with the \texttt{SUBFIND}  algorithm \citep{2001MNRAS.328..726S}. The large volume of \astrid provides a rich galaxy population matching the host galaxy of the observed systems, from which we can conduct a comprehensive search for runaway BHs and star-forming wakes. We perform the search at two redshifts, $z=2$ and $z=1.3$ (the current lowest redshift of the simulation), with the latter being closer in time to the observed streak in VD23 at $z=0.964$.

\textbf{Target galaxies:}
In Figure \ref{fig:z2population}, we show the properties of $z=2$ galaxies in \texttt{ASTRID} and our selection criteria. To match with the properties of the main galaxy in \cite{vanDokkum2023}, we focus on the star-forming (SFR$ > 1\,M_\odot/{\rm yr}$), compact ($r_{\rm half} < 2 {\rm kpc}$), and non-satellite galaxies with masses $5\times 10^9\,\msun < M_{\rm gal} < 5\times 10^{10}\,\msun$ that contain at least 3 BHs with a total BH mass above $10^7\,\msun$. 
We apply the 3-BH criterion because it is necessary to produce a runaway BH. 
After applying these cuts, we narrow down our search to $29888$ (11399) galaxies at $z=1.3$ ($z=2$), and hereafter we refer to these selected galaxies as our ``target galaxies".

\begin{figure*}
    \centering
    \includegraphics[width=1.\textwidth]{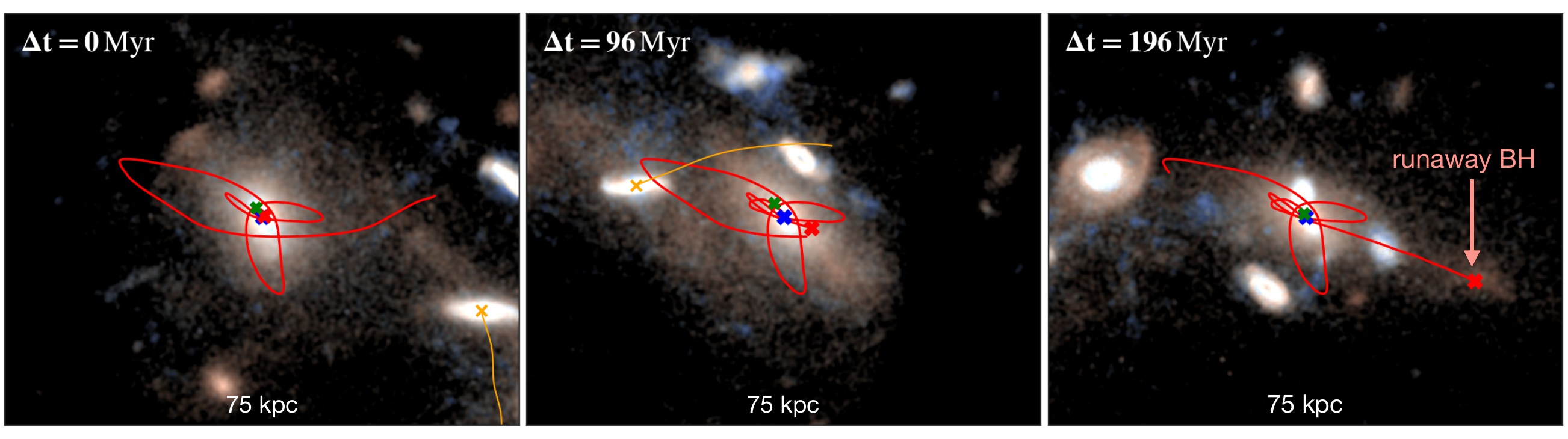}
    \caption{An example runaway-BH involved in an MBH triplet (marked by \textit{red, green, and blue crosses}) identified in \texttt{ASTRID} at $z\sim 1.7$. The \textit{red cross} is the BH that got ejected from the host galaxy with a velocity of $\sim 700\,{\rm km/s}$, due to the complex dynamics within the halo. The host galaxy mass is $M_{\rm gal}=1.3\times 10^{11}\,M_\odot$, and the runaway BH has a mass of $M_{\rm BH}= 1.3\times 10^7\,M_\odot$, and the central BH (\textit{blue cross}) has $M_{\rm BH}= 6.6\times 10^8\,M_\odot$.}
    \label{fig:runaway}
\end{figure*}

\textbf{Runaway BH selection:}
We identify runaway BHs within the target galaxies above (i.e. the orange population in Figure \ref{fig:z2population}). 
To quantitatively define a ``runaway" MBH in these galaxies, we compute the ratio between consecutive apocentric radii between each satellite BH in the galaxy and the central MBH \citep[similar to][]{Hoffman2023}. 
If we see a consistent and sudden increase in orbital size by a factor of more than three in at least three full orbits, then we categorize the MBH as a ``runaway" MBH. Among the $\sim 11399$ target galaxies at $z=2$, we found $\sim 200$ potential runaway BHs that satisfy these criteria. 


\textbf{Linear star-forming wake selection:}
We look for radial linear star-forming features with the following method:
we divide all the young stars (stellar age $< 100\,{\rm Myrs}$) in each halo into 45 radial bins [10, 100] kpc from the central galaxy center. To match a $\sim 50\,{\rm kpc}$ observable linear feature, we select systems with at least $2\times 10^6\msun$ stars in each bin for 25 continuous radial bins. This gives us 514 (435) candidates with potentially observable stellar streaks at $z=1.3$ ($z=2$) respectively, which makes up $2-4\%$ of the target galaxy population (we will refer to these as ``linear feature candidates"). We then visually inspect these potential linear features to select the ones with large signal-to-noise, resulting in $\sim 30$ visually-confirmed linear features at each redshift (note that this number of viable candidates is likely a lower limit as we only look for linearity along the $x$-$y$ and $x$-$z$ planes).


\section{Results}
\label{section:Results}

\subsection{Do runaway BHs produce linear stellar wakes?}
\label{sec:runaway}


\cite{vanDokkum2023} proposed that one of the most likely mechanisms for producing the observed star-forming wake is a runaway BH induced by a three-body encounter. Following this proposal and the works by \cite{Saslaw1972} and \cite{FuenteMarcos2008}, we look for possible associations between linear star-forming features and runaway BHs. 
However, upon an initial search through the target galaxies (orange points in Figure \ref{fig:z2population}), we do not find any visible stellar features along the wakes of the runaway BHs.

As the BH-induced star formation may be more prominent within a denser circum-galactic medium (CGM) than that in our target galaxies, we relax the upper limit on the galaxy mass and did another search among MBH triplets in larger halos (the sample of MBH triplets from \cite{Hoffman2023}). 
In these halos, the runaway BHs also tend to be more massive, and the formation of a stellar wake is less affected by the resolution limit. 
Figure \ref{fig:runaway} shows an example runaway BH candidate within a MBH triplet, found in a galaxy with mass $1.3\times 10^{11}\,M_\odot$. 
In this system, a BH with mass $1.3\times 10^7\,M_\odot$ is ejected from the host galaxy $\sim 100\,{\rm Myrs}$ after the initial formation of the MBH triplet at a speed of $\sim 700\,{\rm km/s}$, and it traverses a distance of $\sim 45\,{\rm kpc}$ from the galaxy center in another $\sim 100\,{\rm Myrs}$. 
Even in this case, we still do not see a star-forming signature associated with the wake of the runaway BH: we found at most $<10^6M_\odot$ total newly formed stars within $10\,{\rm kpc}$ of the ejected BH.

We note a few limitations of the simulation potentially responsible for this null result: the resolution of \texttt{ASTRID} cannot fully resolve the ejection due to three-body encounters, and BHs that ``run away" are mostly disrupted by a blob of incoming stars/gas. 
This implies that the maximum speed reached by any of our runaway BH candidates is $\sim 800\,{\rm km/s}$ and decreases to $\sim 100\,{\rm km/s}$ at the target separation of $\sim 50\,{\rm kpc}$, barely reaching the speed to shock the surrounding gas \citep[e.g.][]{Dopita1996, Allen2008}. 
Also, the simulation may lack the mass resolution to resolve gas shocked by a BH: the target runaway BH ($M_{\rm BH} \sim 10^7M_\odot$) is a few times larger than the gas particle mass, and so the gas over-density drawn by the BH may not be directly resolvable. A follow-up study would need to deploy higher-resolution simulations, to detect BHs ejected at a higher velocity in environments similar to our target galaxies to evaluate whether a similar star-forming wake can be produced.


\subsection{Stellar wakes associated with fly-by galaxy encounters}
\label{sec:feature}

\begin{figure*}
    \centering
    \includegraphics[width=1.\textwidth]{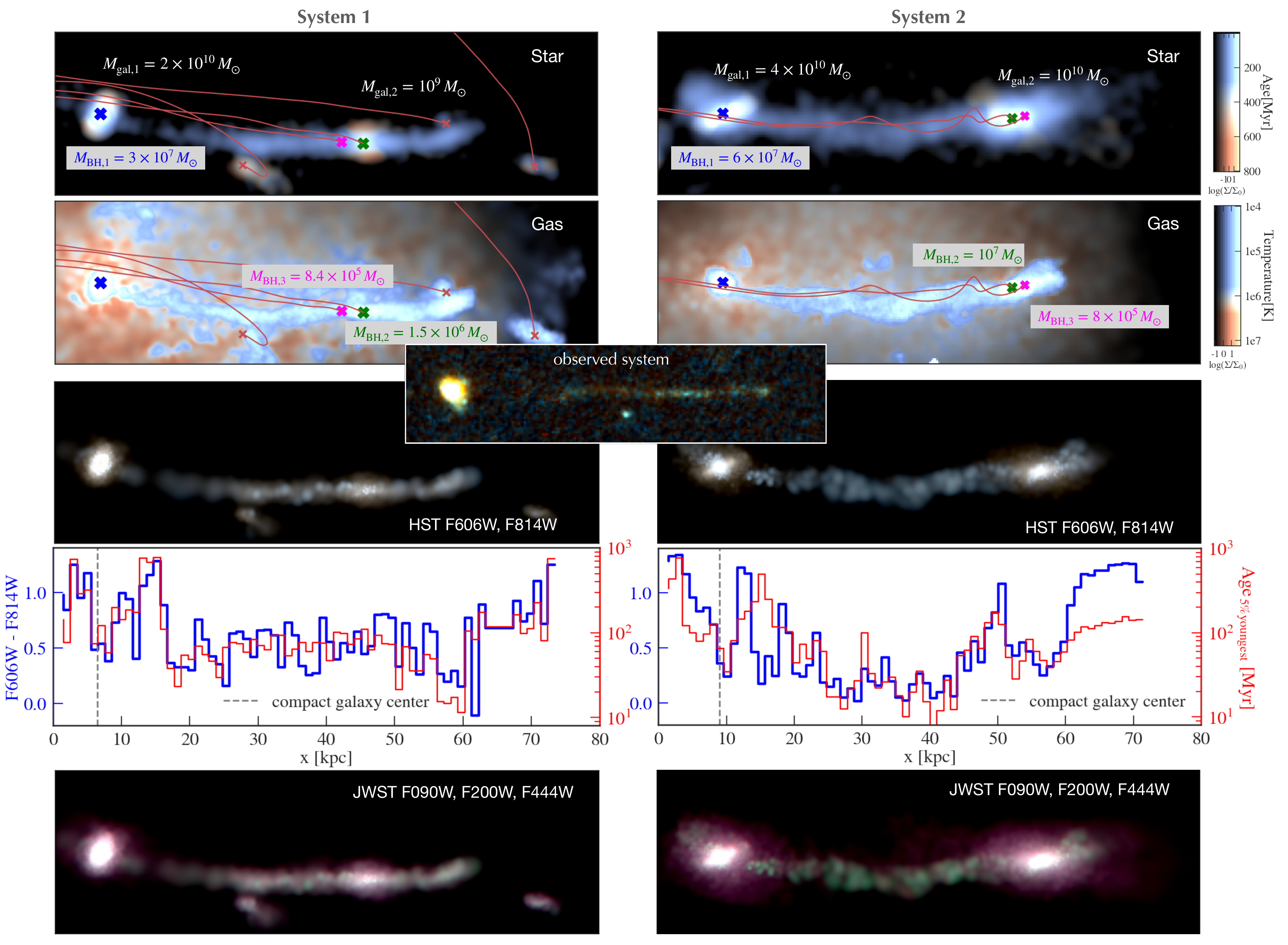}
    \caption{\textbf{Row 1/2:} The stars (first row, color-coded by age) and gas (second row, color-coded by temperature) associated with the linear stellar streaks at $z\sim 2$ (\textit{left}) and $z\sim 1$ (\textit{right}). 
    The linear features extend to $\sim 50\,{\rm kpc}$ from the compact galaxy on the left of each image. 
    \textbf{Row 3:} Mock image of the systems in the HST F606W and F814W filters.
    \textbf{Row 4:} F606W - F814W color of the stellar population, modeled with with the \texttt{Python-FSPS} stellar population model \citep{Conroy2010}, and the mean age of the $5\%$ youngest stars.
    \textbf{Row 5:} Mock image of the systems in the JWST filters will reveal the older stellar population along the wake.
 }
    \label{fig:images}
\end{figure*}

Although we do not find visible star-forming wakes associated with the passage of runaway BHs, we do find a few tens of star-forming wakes in the simulation following the method detailed in Section \ref{section2:Method}. 
The natural question to ask is then: what mechanisms, if not runaway BHs, produce these wakes? 
By inspecting the simulation counterparts to the observed systems selected in Section \ref{sec:selection}, we find that most star-forming streaks originate from the fly-by encounter of a recently merged, dual-BH young galaxy with a more massive compact galaxy. 
We present some representative cases of such systems and their evolution in this section.

Figure \ref{fig:images} shows star and gas visualizations of two systems with linear star-forming features, and the mock images of the systems seen through the different filters of HST and JWST. 
In both cases, the young stars extend linearly to more than 50 kpc away from the main, compact galaxy (Galaxy 1, on the left end of each image). The young galaxy (Galaxy 2, on the right) is embedded in a stream of cold, dense gas with $T\sim 10^4-10^5\,{\rm K}$ clearly distinguishable from the background hot gas in the surrounding medium (second row), along which we see ongoing star formation and young stars formed within a few tens Myrs (fourth row).

Figure \ref{fig:images} also shows the mock HST and JWST observations, created by assigning spectral energy distributions (SEDs) to each star particle according to the Binary Population and Spectral Population Synthesis model \citep[BPASS v2.2.1;][]{Stanway2018}. 
These SEDs are convolved with the filter transmission functions associated with the chosen filters.
We smooth the stars with an SPH kernel and make 2D projections with pixel sizes matching the sensor for HST and JWST (0.049" for HST ACS, and 0.031"/0.063" for JWST NIRCAM short/long wavelength).

The stellar age along the wake follows a bi-modal distribution: half of the stars are relatively old with ages $\sim 1\,{\rm Gyr}$, and half are formed within $<100\,{\rm Myrs}$ during the emergence of the streak. 
The young stars form the linear feature when captured by the HST F606W and F814W bands (third row), while the older stars are better seen with JWST at longer wavelengths (fifth row). The redder stars may be a distinctive signature of the stripped galaxy scenario, which are unlikely to be found in the runaway-BH case.

\subsection{Evolution of the linear feature}
\begin{figure*}
    \centering
{
\includegraphics[width=0.87\textwidth]{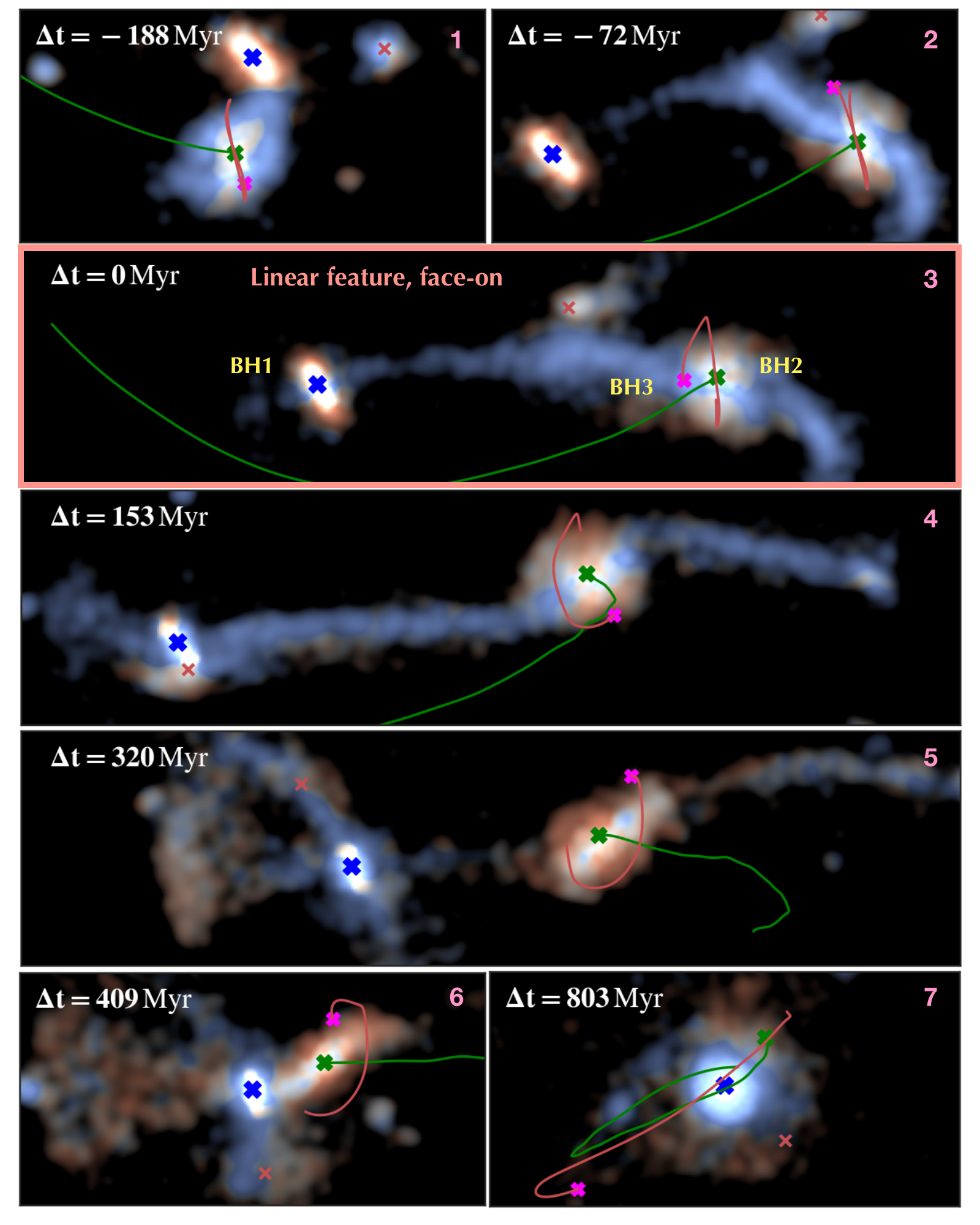}
}
    \caption{ Time evolution of the two galaxies involved in producing the linear stellar feature (projected face-on), and the associated BHs (central BH in the Galaxy 1 in \textit{blue}, two BHs in the tidally disrupted galaxy in \textit{green and magenta}, and the another BH nearby in \textit{red}). 
    The green line traces the trajectory of BH2 relative to BH1, and the red line shows the orbit of BH3 relative to BH2.
    }
    \label{fig:evolution}
\end{figure*}

\begin{figure}[h]
    \centering
    \includegraphics[width=0.48\textwidth]{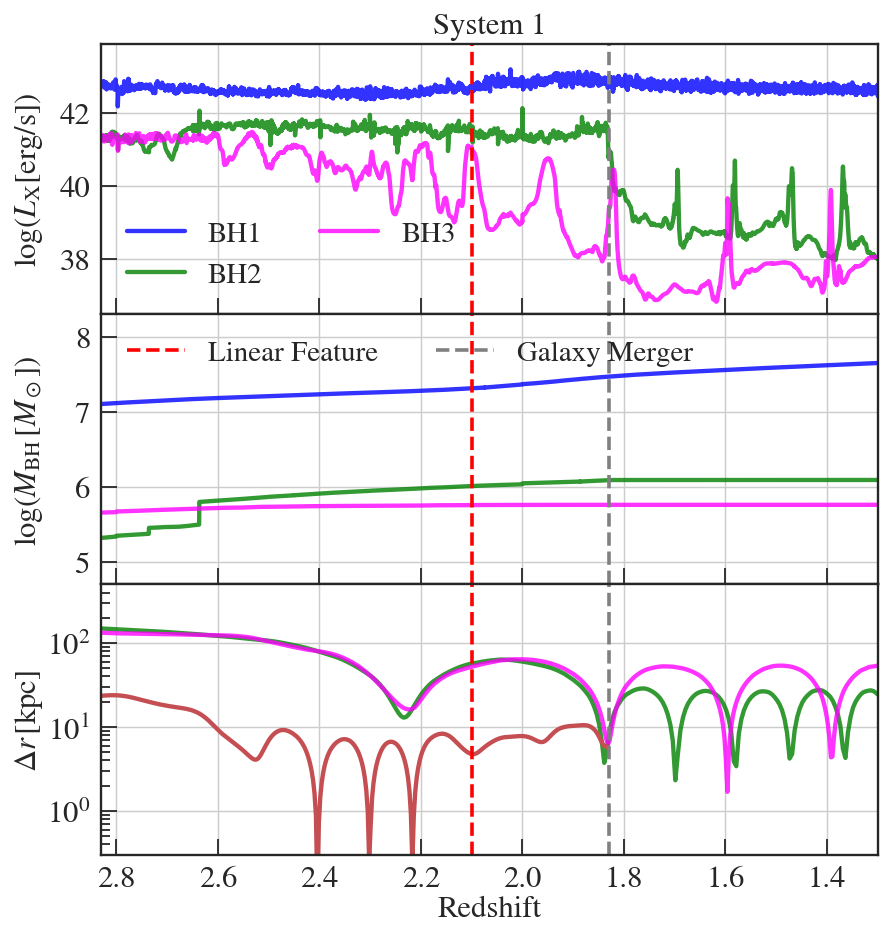}
    \includegraphics[width=0.48\textwidth]{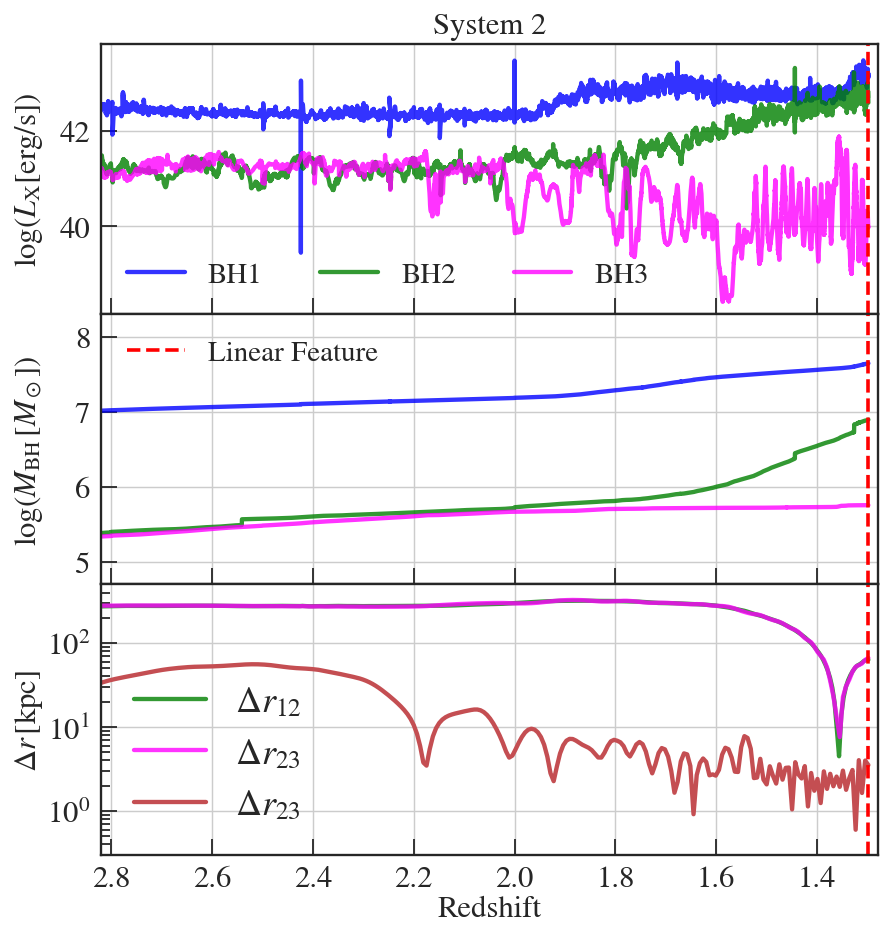}
    \caption{Properties of the three BHs along the linear stellar feature. The top two panels show the X-ray luminosities and BH masses of the three BHs. The bottom panel shows the distance between the three BHs.}
    \label{fig:bhs}
\end{figure}

To better understand the formation of the star-forming streak, we trace System 1 across a Gyr before and after the time when the linear feature is most prominent. Figure \ref{fig:evolution} shows the time evolution of the two galaxies involved in the linear stellar feature in a face-on view. 
The feature is a by-product of a fly-by encounter between two galaxies at a speed of $580\,{\rm km/s}$. 
The galaxies barely touch each other during their first passage (frame 1), after which Galaxy 1 remains unperturbed while Galaxy 2 develops elongated tidal arms on both sides. 
The feature formed around $150\,{\rm Myrs}$ after the initial passage between the two galaxies, and lasts for $\sim 200\,{\rm Myrs}$ (frame 2-4), after which the stars along the streak become old (frame 5), and the two galaxies experience a head-on collision (frame 6-7).

Our simulated linear features are extreme cases of the tidal arms/tails found in previous numerical studies of galaxy mergers \citep[e.g.][]{Toomre1972, Barnes1992, Ploeckinger2018}. 
Idealized simulations have shown that within these tails accumulations of gas and stars can be found \citep[e.g.][]{Barnes1992, Baumschlager2019, Shin2020}.
These clumps harbor molecular gas which provides a reservoir for new stars \citep[e.g.][]{Mirabel1991, Mirabel1992}, and could be the origin of the blue knots along the VD23 system.
Upon the close inspection of the $\sim 50$ visually-confirmed linear features, we find that the long tails are usually associated with dual-BH galaxies.
$70\%$ of these tails contain close BH pairs, indicating a recent galaxy merger within the tail.
Isolated galaxy simulations may not produce features like System 1 and System 2 if it takes two consecutive galaxy mergers to produce very long star-forming tails.

We also show in Figure \ref{fig:evolution} the trajectories of the BHs along the tail. 
BH3 has come within Galaxy 2 in a previous galaxy merger a few hundred Myrs before the wake formation.
Then it keeps orbiting around BH2 during and after the production of the stellar wake, until the final merger between Galaxy 1 and Galaxy 2 (frame 7), after which both BH2 and BH3 ended up in wide orbits around BH1.
We will examine the detailed evolution of the triple-BH system in the next section.


\subsection{BHs along the linear feature}
\label{sec:bh}
A common feature shared by our linear feature candidates is a dual-BH system along the streak merging into a central BH. 
Such MBH triplets are indicators of consecutive galaxy mergers producing the stellar streak, and if active, they may be also responsible for triggering the OIII emission along the streak. Here we examine the evolution of BHs along the linear features and their AGN activity.

Figure \ref{fig:bhs} shows the time evolution of the X-ray luminosity, BH mass, and the orbits of the three BHs associated with the merging galaxies in System 1 and System 2. 
In both cases, the BH in the compact galaxy is fairly luminous, maintaining an X-ray luminosity of nearly  $10^{43}\,{\rm erg/s}$ for more than a Gyr. 
The second brightest BH (BH2) embedded along the linear feature also has a relatively high X-ray emission between $10^{42}-10^{43}\,{\rm erg/s}$. 
If the observed wake is produced by the scenario we show here, then BH2 is responsible for triggering the OIII emission along the wake. As mentioned in VD23, the amount of OIII emission seen would correspond to an AGN with $L_X>10^{43}\,{\rm erg/s}$ according to the scaling relation in \cite{Ueda2015}. Our BH2s are slightly fainter for a $1.9 \times 10^{41}\,{\rm erg/s}$ OIII luminosity, but still fall well within the \cite{Ueda2015} relation with the scatter and are capable of triggering the observed OIII signature.

The detectability of the third BH may depend on its orbits within Galaxy 2, but it is possible that BH3 was captured at the pericenter of the orbit and at it is brightest when the linear feature is produced (e.g. the case in System 1). In this (optimistic) case, follow-up observations may simultaneously see an MBH triplet associated with the stellar wake.


\section{Conclusions and Discussion}
\label{section:conclusion}

In this work, we identify runaway BHs ejected at $\sim 700\,{\rm km/s}$ from host galaxies, as well as linear star-forming features extending $>50\,{\rm kpc}$ from compact galaxies in the \texttt{ASTRID} simulation at $z=1\sim2$. 
We look for the association between the two types of systems following the proposal in VD23, and find that runaway BHs do not produce visible star-forming wakes in the simulation.

However, we find that linear star-forming features in fly-by galaxy encounters resemble the recent observation by \cite{vanDokkum2023}, with very bright, thin, and straight young stars when seen in the HST F606W/F814W filters. 
Such linear features potentially exist among $\sim 0.1\%$ compact, star-forming galaxies with $5\times 10^9\,\msun < M_{\rm gal} < 5\times 10^{10}\,\msun$ and multiple BHs, and the most prominent linear features can last for $100-200\,{\rm Myrs}$.

We examine two representative systems with strong linear features in detail, both with stellar masses between $10^{10}-5\times 10^{10}\,M_\odot$  and SFRs between $10-50 M_\odot/{\rm yr}$. Among the BHs associated with the two galaxies producing the linear feature, the central BH has a bright X-ray emission with $10^{42} \,{\rm erg/s}< L_X < 10^{43}\,{\rm erg/s}$, which could potentially be seen by Chandra. 
The brightest BH along the feature emits at $10^{41} \,{\rm erg/s}< L_X < 10^{43}\,{\rm erg/s}$, also with a chance to be detected in the X-rays. 

The production of the linear feature involves consecutive encounters between three galaxies. 
The remnant of the first galaxy merger ($\sim 10^9-10^{10}\,M_\odot$ and star-forming) goes through a fly-by encounter with a more massive, compact galaxy, during which its surrounding gas gets tidally stripped, leaving a trace of star-forming gas and young stars. 
$\sim 70\%$ of the visually-confirmed systems with linear features harbor a dual-BH star-forming galaxy along the feature, supporting the three-galaxy encounter scenario. Our findings indicate that these linear features may offer a robust signature for finding BH binaries and multiple-BH systems, but further work is required to better establish the connection between linear long tidal tails and binary-hosting galaxies. There have been catalogs of long tidal tails in the COSMOS field \citep[e.g.][]{Civano+2010, Wen2016, Ren2020}, out of which two galaxy nuclei can be clearly identified in some extended linear systems. 

Finally, although our results show that only tidal features during fly-by galaxy mergers can produce these recently observed features, we acknowledge that our resolution limit leaves room for the possibility of a runaway-BH-induced star-forming wake. Due to the similarity with many features of the observed signatures (star-forming gas, OIII emission, and a population of young stars) shared by these two scenarios, follow-up studies are needed to potentially distinguish between them.

Follow-up observations using longer-wavelength bands with JWST can tell whether there exists an underlying old stellar population along the linear feature. The lack of old stars will greatly support the runaway BH origin of the feature. Meanwhile, higher-resolution simulations can study the triggering of such dense star formation by runaway BHs, and the mass/velocity of the BH required. In both formation channels of the linear feature, follow-up observations may be able to reveal multiple MBH signatures along the feature, and such systems are very likely the sites for complex three-body encounters and mergers between MBHs \citep[e.g.][]{Hoffman2007,Bonetti2016,Mannerkoski2021}.

\section*{Acknowledgements}

We thank Mohit Bhardwaj, Pieter van Dokkum, Charlie Conroy and Qian Yang for their helpful discussions. 
\texttt{ASTRID} was run on the Frontera facility at the Texas Advanced Computing Center.
TDM and RACC acknowledge funding from the NSF AI Institute: Physics of the Future, NSF PHY-2020295, NASA ATP NNX17AK56G, and NASA ATP 80NSSC18K101.
TDM acknowledges additional support from  NSF ACI-1614853, NSF AST-1616168, NASA ATP 19-ATP19-0084, and NASA ATP 80NSSC20K0519, and RACC from NSF AST-1909193.
SB acknowledges funding supported by NASA-80NSSC22K1897.

\section*{Data Availability}
The code to reproduce the simulation is available at \url{https://github.com/MP-Gadget/MP-Gadget}, and continues to be developed.
Part of the \astrid snapshots are available at \url{https://astrid-portal.psc.edu/}.

\bibliography{main}{}
\bibliographystyle{aasjournal}

\end{document}